\documentclass[12pt,preprint]{aastex}

\slugcomment{To appear in ApJ Letters}

\shorttitle{QUEST RR Lyrae Survey}
\shortauthors{Vivas et al.}

\begin{document}

\title{The QUEST RR Lyrae Survey: Confirmation of the Clump at 50
kpc and Other Over-Densities in the Outer Halo}

\author{A. K. Vivas\altaffilmark{1}, R. Zinn\altaffilmark{1},
P. Andrews\altaffilmark{2}, C. Bailyn\altaffilmark{1},
C. Baltay\altaffilmark{2}, P. Coppi\altaffilmark{1,2},
N. Ellman\altaffilmark{2}, T. Girard\altaffilmark{1},
D. Rabinowitz\altaffilmark{2}, B. Schaefer\altaffilmark{3}, 
J. Shin\altaffilmark{2}, J. Snyder\altaffilmark{2,1},
S. Sofia\altaffilmark{1}, W. van Altena\altaffilmark{1},
C. Abad\altaffilmark{4}, A. Bongiovanni\altaffilmark{4},
C. Brice\~no\altaffilmark{4}, G. Bruzual\altaffilmark{4}, 
F. Della Prugna\altaffilmark{4}, D. Herrera\altaffilmark{4}, 
G. Magris\altaffilmark{4}, J. Mateu\altaffilmark{4}, 
R. Pacheco\altaffilmark{4}, Ge. S\'anchez\altaffilmark{4}, 
Gu. S\'anchez\altaffilmark{4}, H. Schenner\altaffilmark{4}, 
J. Stock\altaffilmark{4}, B. Vicente\altaffilmark{4},
K. Vieira\altaffilmark{4}, I. Ferr{\'\i}n\altaffilmark{5}, 
J. Hernandez\altaffilmark{5}, M. Gebhard\altaffilmark{6},
R. Honeycutt\altaffilmark{6}, S. Mufson\altaffilmark{6},
J. Musser\altaffilmark{6}, A. Rengstorf\altaffilmark{6}}

\altaffiltext{1}{Yale University, Astronomy Dept. New Haven, CT}
\altaffiltext{2}{Yale University, Physics Dept. New Haven, CT}
\altaffiltext{3}{University of Texas, Austin TX}
\altaffiltext{4}{Centro de Investigaciones de Astronom{\'\i}a (CIDA),
M\'erida, Venezuela}
\altaffiltext{5}{Universidad de Los Andes, M\'erida, Venezuela}
\altaffiltext{6}{Indiana University, Bloomington IN}

\begin{abstract}
We have measured the periods and light curves of 148 RR Lyrae
variables from V=13.5 to 19.7 from the first 100 deg$^2$ of the QUEST
RR Lyrae survey.  Approximately 55\% of these stars belong to the
clump of stars detected earlier by the Sloan Digital Sky
Survey. According to our measurements, this feature has $\sim$10 times
the background density of halo stars, spans at least $37.5^\circ$ by
$3.5^\circ$ in $\alpha$ and $\delta$ ($\ge30$ by $\ge3$kpc), lies
$\sim50$ kpc from the Sun, and has a depth along the line of sight of
$\sim5$ kpc ($1\sigma$). These properties are consistent with the
recent models that suggest it is a tidal stream from the Sgr dSph
galaxy. The mean period of the type ab variables, $0.58^d$, is also
consistent. In addition, we have found two smaller over-densities in
the halo, one of which may be related to the globular cluster Pal 5.
\end{abstract}


\section{Introduction}
A well-supported working hypothesis for the formation of the galactic
halo is that its stars and globular clusters were torn from smaller
galaxies as they underwent tidal destruction on the way to merging
with the Milky Way (see review by Bland-Hawthorn and Freeman 2000).
While many features of the halo support this picture, the most direct
evidence is provided by the Sagittarius dwarf spheroidal (dSph) galaxy
which is now in the process of tidal destruction in the halo (Ibata,
Gilmore \& Irwin 1994, Mateo et al. 1996) .   Models of the demise of
the Sgr dSph and of other small galaxies have shown that streams of
stars from merger events should persist for billions of years in the
halo (Johnston 1998, Helmi \& White 1999). 
Until very recently, these streams escaped detection because the old
surveys of the halo were too limited in sky coverage or in
depth. While a few moving groups of high velocity stars have been
known for a long time (e.g. Eggen 1979), these could be explained by
the disruption of star clusters (Eggen 1996), and the density contours
of the halo were thought to be smooth on scales of kiloparsecs.
The new surveys that map the
densities of halo stars over larger areas and to greater depths, and
ones that detect moving groups of halo stars have revealed large
substructures in both density and velocity space, as predicted by the
merger hypothesis (Majewski, Munn \& Hawley 1994, Helmi et al. 1999,
Morrison et al. 2000, Ibata et al. 2001a).  
One of the most significant results of these
surveys was the discovery of a large clump of candidate RR Lyrae
variables (RRs) and A type stars by the Sloan Digital Sky Survey (SDSS)
(Ivezic et al. 2000, Yanny et al. 2000). Ivezic et al. and Yanny et
al.  suggested that this clump is part of the tidal tail of the Sgr
dSph, which is supported by the more recent modeling of the Sgr stream
by Ibata et al. (2001b), Helmi \& White (2001) and
Mart{\'\i}nez-Delgado et al. (2001).

We report here the first results of the QUEST survey for RRs, which
will ultimately survey about 700 deg$^2$ of the sky down to a
limiting magnitude of V$\sim$19.7.  By coincidence, the first part
that has been completed includes a large fraction of the region
searched earlier by the SDSS for RRs.   Unlike the SDSS, which
consisted of only two epochs of observations separated by 1.99 days,
we have observations at 20 to 30 epochs separated by hours to more
than one year.  These observations allow us to detect many more RRs
than the SDSS and to determine the periods and the light curves of the
stars, which ensures that they are indeed RRs and not another type of
variable. Also, we determine the mean magnitude of the stars from many
measurements which reduces the errors in their distances.

Our results confirm the major conclusion of the SDSS that a
large clump of old stars exists in the halo at $\sim$50 kpc from the
Sun and show that this clump extends beyond the SDSS region.  We also
find evidence for two smaller density enhancements, one of which may
be related to the globular cluster Pal 5.

\section{Observations and Data Reduction}
The observations for this survey were made with the 1m Schmidt
telescope at Llano del Hato Observatory  in Venezuela and the CCD
camera built by the QUEST (Quasar Equatorial Survey Team)
collaboration (Snyder 1998).  The camera has 16 $2048\times 2048$ CCDs
set in a $4\times 4$ array.  It has a $2.3^\circ\times 2.3^\circ$
field of view, and pixel size of 15 microns corresponds to a scale
of $1^{\prime\prime}.02$ per pixel.  The camera is operated in
drift-scan mode, which generates a continuous strip of the sky at a
rate of 34.5 deg$^2$/hr.  Each row of 4 CCDs is fitted with a
different filter. Because V observations were obtained every
night, they were chosen for the search for RRs. The limiting
magnitude of the survey is $\sim$19.7 in V (S/N=10), which was set by
the drift time of a star across a chip (140 sec at $\delta$=0) and the
seeing, which varied from 1.9 to 3.2 arcsec.  Saturation of the CCDs
occurred at V=13.5.  We obtained between 20 and 30 scans of the region
$\alpha$=13$^h$ to 16$^h$, $\delta$=$-2^\circ 20^\prime$ to $+0^\circ
02^\prime$ (100 deg$^2$)  during Feb 1999 to Apr 2000. Because of the
defects of the CCDs or high levels of cloudiness, a small fraction of
the stars could not be measured more than 12 times.

The standard QUEST software performed bias subtraction, flatfielding
and aperture photometry in each of the 4 filter bands that an object
was observed. The instrumental magnitudes were normalized to a
reference scan in order to remove variable atmospheric extinction due
to differences in air mass and transparency.  This normalization to
reference magnitudes was made in bins of 0.25 deg of right
ascension ($\sim$1 minute of time in drift scan mode).  Each bin
contains typically 1000 stars, which produces a robust estimate of the
zero point difference between the scan and the reference scan.  The
random errors of the differential magnitudes amount to 0.02 for V$\le$
17, but climb to 0.12 at V=19.7 as photon statistics dominate the
noise. Roughly 40 secondary standard stars were setup with the 1m YALO
telescope at CTIO. The zero point error in V is $\sim0.03$ mag.

To find the variable stars, we used a $\chi^2$ test (e.g. Saha \&
Hoessel, 1990) on the normalized instrumental magnitudes.  If the
probability that the observed variation of a star is caused by errors
alone is very small ($<$0.01), then the star is labeled as variable.
The RRs  were identified by first reducing the list of variable stars
to those with amplitudes between 0.25 and 1.5 magnitudes in V and with
V-R colors $<$ 0.55.  We then isolated stars with periods in the range
of 0.2 to 1.0 days.  The phased light curves were examined by eye, and
RRs of type ab were easily recognized by the light-curve shapes.
Although this method also detected several type c RRs, which have
lower amplitudes and more sinusoidal light-curves than the type ab, it
is not optimum for finding them.  We will search our data later for
type c, which are less numerous than the type ab variables.  The
quality of the light curves that are produced by the data is
illustrated in Figure 1.

The completeness of the survey is a complex problem because the number
of observations of each star depends on position on the sky and
magnitude. Using Monte Carlo simulations we estimate that the survey
is $\sim$80-100\% complete for $V<18.5$ and $\sim$55-78\% complete at
the faint end of the survey. A more detailed description of these
simulations and the techniques of the survey are given in Vivas et
al. (2001).

\section{Results}
The survey found a total of 135 RRs of type ab and 13 of type c.
The distributions of these stars on the sky and in magnitude as a
function of right ascension are shown in Figure 2.  This figure also
illustrates the region of overlap with the SDSS.  The magnitudes were
corrected for interstellar extinction using the maps of Schlegel et
al. (1998).

In the region of overlap, the SDSS identified 59 stars as RRs.  Our
survey recovered 50 of these stars or 85\%.  The remaining 9 stars
were either so faint that we had too few observations to calculate
periods or were missed by our software because they lie on bad columns
in the reference scan.  Of the 50 stars for which we had adequate
observations, 34 were confirmed to be RRs of type ab and 4 were
confirmed as type c.  We suspect that the remaining 12 stars are not
RRs or ones whose light curves are strongly affected by the Blazhko
effect. However, we found in addition many more RRs. We found a
total of 90 RRs in this region, of which only  42\% were  identified by
the two epoch SDSS survey. This is consistent with the completeness
estimate made by Ivezic et al. 2000.

The plot of the mean magnitudes against right ascension  in the lower
diagram of Figure 2 shows a very conspicuous clustering in $\langle
V_0 \rangle$ between $\sim$19.0 to 19.5.   This is the clump of stars
that the SDSS (Ivezic et al. 2000, Yanny et al. 2000)
discovered earlier.  Our survey confirms the existence of this feature
of the galactic halo, which is conclusive evidence that the density
contours of the halo are not smooth (see also Fig. 4), and provides a detailed
description of its properties.

The over-density of the clump with respect to the background of halo
stars appears to be about a factor of 10 (see below).  Consequently,
little error is probably introduced if we consider all of the 84 stars
with $\langle V_0 \rangle \ge 18.4$ to be clump members when
estimating its dimensions.  The clump extends from the western limit
of our survey at 13$^h$ to at least 15.5$^h$, a span of 37.5$^\circ$ (see
Fig. 2).  Between 13$^h$ and $\sim$14.2$^h$, the faint limit of the clump
appears to be significantly brighter than the limit of our survey.
This is less clear from 14.2$^h$ to 15.5$^h$, and the apparent disappearance
of the clump between $\sim$15.5$^h$ and 16$^h$ may be caused by it moving
fainter than our limit.  Our data from 13$^h$ to 14$^h$ indicate that the
standard deviation in $\langle V_0 \rangle$ is only 0.26 mag, which we
show below is evidence for a relatively small depth along the line of
sight.  This is consistent with the SDSS, which did not find
any RRs at $r'>20$.  From 13$^h$ to 15.5$^h$, clump stars are found at all
$\delta$ in our survey.  If we consider the additional area surveyed
by the SDSS, the clump extends at least 3.5$^\circ$ in $\delta$.  The
highest concentration of stars occurs at $\alpha$ = 14.6$^h$, where there
are as many as 2 RRs per deg$^2$.  There is no obvious concentration
in $\delta$, although there is a slight decrease in RRs from north to
south.

To calculate the distances to the RRs, we have assumed that they have
an absolute visual magnitude ($M_V$) of +0.56. This value is
appropriate for a stellar population with [Fe/H] = -1.6, which is
typical of the halo,  and a horizontal branch (HB)  morphology that
produces a large population of RRs (Demarque et al. 2000). It is near
the middle of the range given for this [Fe/H] by several observational
studies (see Vivas et al. 2001 for more discussion).
With this value of $M_V$, the distance to the clump is approximately
50 kpc, and its minimum dimensions of 37.5$^\circ$ and 3.5$^\circ$ in
$\alpha$ and $\delta$,  respectively, translate into $\ge 30$ by $\ge 3$
kpc.  The dispersion in $\langle V_0 \rangle$  (0.26 mag, see above)
is due to the observational errors in $\langle V \rangle$ ($\sim$0.08
mag), the errors in the correction for interstellar extinction
($\sim$0.06), the dispersion in the distance modulus, and two effects
related to the properties of HB stars, the dispersion in RR absolute
magnitude caused by evolution ($\sim$0.08 mag from the CMDs of
globular clusters, e.g. Table 13 in Sandage 1990) and by the
dispersion in [Fe/H] in the stellar population.  Taking $\sigma$[Fe/H]
= 0.5 as a value typical of dwarf galaxies (including Sgr, Mateo 1998) 
and the Demarque et al. (2000) $M_V$-[Fe/H]
relation, we estimate a dispersion of $\sim$0.1 from this source.
Subtracting in quadrature the observational errors and these other
estimates from the observed dispersion in $\langle V_0 \rangle$,  we
find a dispersion of 0.2 mag in the distance modulus.  This suggests that
the depth along the line of sight has a dispersion of $\sim$5 kpc.
The clump is long, $\ge 30$ kpc, and thin, $\sim$5 kpc, but until more is
known about its extent in $\delta$, we do not know whether it
resembles a sheet or a ribbon.

To calculate the number density of RRs, we divided the survey into 8
equal strips in $\alpha$ and followed Saha's (1985) procedure in each
strip.  After obtaining the densities along the lines of sight, we
transformed from heliocentric to galactocentric coordinates and
averaged the densities in bins of equal size in log galactocentric
distance (R).  These densities are plotted against log R in the top
diagram of Figure 3, where the line was fitted to the 9 points that
define a lower envelope to the densities.  The slope of this line
(-3.0$\pm$0.2) is consistent with previous determinations (Wetterer \&
McGraw 1996, Ivezic et al. 2000) of the density fall-off in the
halo\footnote{The referee has pointed out that a fit to all the data
in the top panel of Fig 3 would yield a slope near -2, with an
under-density between 25 and 35 kpc, and that this slope is
interesting because it is similar to estimates of the slope of the
dark matter halo. While we too find this interesting, we are reluctant
to draw conclusions from it at this time because in directions away
from the clump of RRs at R$\sim$45 kpc, the slope definitely
steepens to $\sim -3$.}.
The clump of RRs produces the over-density at $R>40$ kpc, where it is
roughly 10 times the density predicted by the line.

In this figure there is evidence for another over-density of RRs from
R$\sim$16 to 23 kpc.  Unlike the distant clump, which is seen in all
but the last one of the strips in $\alpha$, these over-densities are
visible in only two strips and appear to be two separate clusterings.
The lower diagram in Figure 3 shows separately the density gradients
in the two strips.  The most prominent one, which occurs at R$\sim$17
kpc is probably related to the globular cluster Pal 5, which lies
within the area we surveyed on the sky.  Although our software is not
designed to work in crowded regions, we recovered 2 of the 5
known RR Lyrae variables in Pal 5 (all type c, Sawyer Hogg 1973).
These stars were excluded before computing the densities that are
plotted in Figure 3.  The over-density is produced by 6 other RRs, two
of which lie within the tidal radius of Pal 5 and four more distant
ones that are within a projected distance from the cluster of 0.6 kpc.
Since other evidence suggests that Pal 5 is being disrupted by the
tidal field of the Milky Way (Odenkirchen et al. 2001),
we suspect that these variables had their origin in
Pal 5.  The other over-density is produced by 5 stars that have very
similar values of R and lie in the strip of $\alpha$ from 13.0 to
13.375h.  Four of these stars lie within $1^\circ$ of each other,
which is suggestive of a real clustering (see also Figs. 2 \& 4).
Observations will be obtained soon to see if the radial velocities of
these stars and the ones near Pal 5 are consistent with physical
associations.

\section{Discussion}
The long and relatively thin clump of RRs at R$\sim$50 kpc resembles
the tidal streams that have been predicted by models of the merging of
small satellite galaxies with the Milky Way.  After the discovery of this
clump by the SDSS,
Ibata et al. (2001b), Helmi \& White (2001) and Mart{\'\i}nez-Delgado
et al. (2001) have shown that it may be explained as a tidal stream
from the Sgr dSph galaxy.  These models predict that a quite wide stream
crosses the area that we observed at an angle of $\sim$35$^\circ$ with
respect to $\alpha$. They also estimate that
the RRs in this part of the stream must have a magnitude V$\sim$19.5,
consistent with our observations.
Our data show that the mean period of the type ab RRs in the
clump is 0.58
days, similar to the value found in the central part of the Sgr dSph
(Mateo 1996).  This is not strong evidence for an association, for the
RRs outside the clump are only marginally different ($\langle P
\rangle$ = 0.56 day).

Figure 4 is a three dimensional representation of the Milky Way in
which we have plotted the RRs in our survey, Pal 5 and the Sgr dSph
galaxy.  This diagram illustrates the uneven distribution of the RRs
and also the very large distance separating the clump of RRs and the
Sgr dSph galaxy.  The idea that the halo has smooth contours in
density that vary systematically is clearly incompatible with this
diagram.  Even if the clump is unrelated to the Sgr dSph galaxy, it
has the features of a merger remnant and is therefore strong evidence
in favor of the hypothesis that the stars and also the globular
clusters of the outer halo are the debris left over from the accretion
of small galaxies.

\acknowledgments

The Llano del Hato Observatory is operated by CIDA for the Consejo
Nacional de Investigaciones Cient{\'\i}ficas y Tecnol\'ogicas and the
Ministerio de Ciencia y Tecnolog{\'\i}a. This research was partially
supported by the National Science Foundation under grant AST-0098428.

\clearpage
\begin{figure}
\plotone{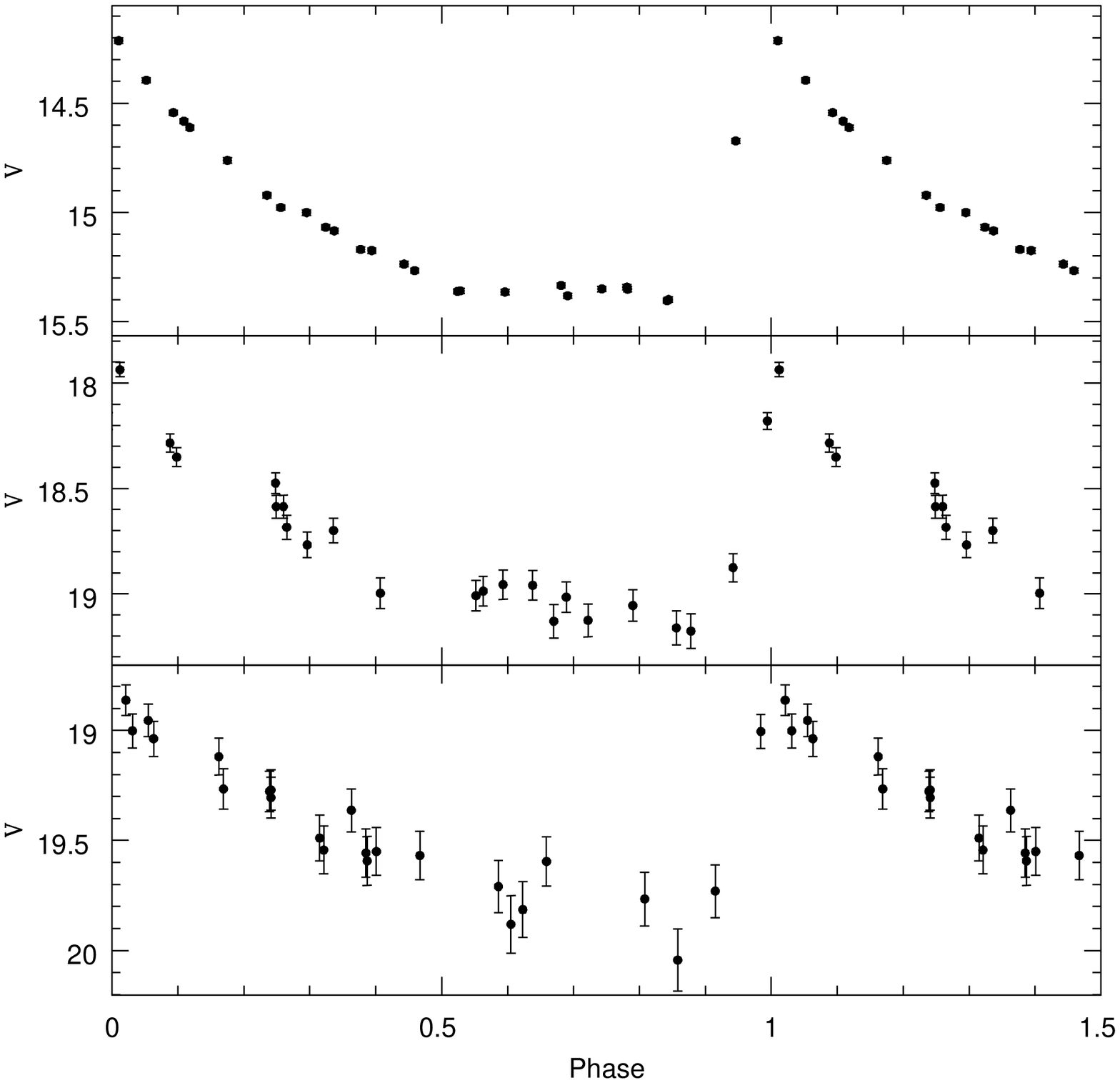}
\caption{Phased light curves of 3 RRs of different magnitudes. The top
diagram is the light curve of a star
near the bright limit of our survey, while the other two are light
curves of stars near the faint limit.  As expected, near the faint
limit the photometric errors are large, but are nonetheless much smaller
than the amplitudes of the type ab variables.
\label{fig1}}
\end{figure}

\clearpage
\begin{figure}
\plotone{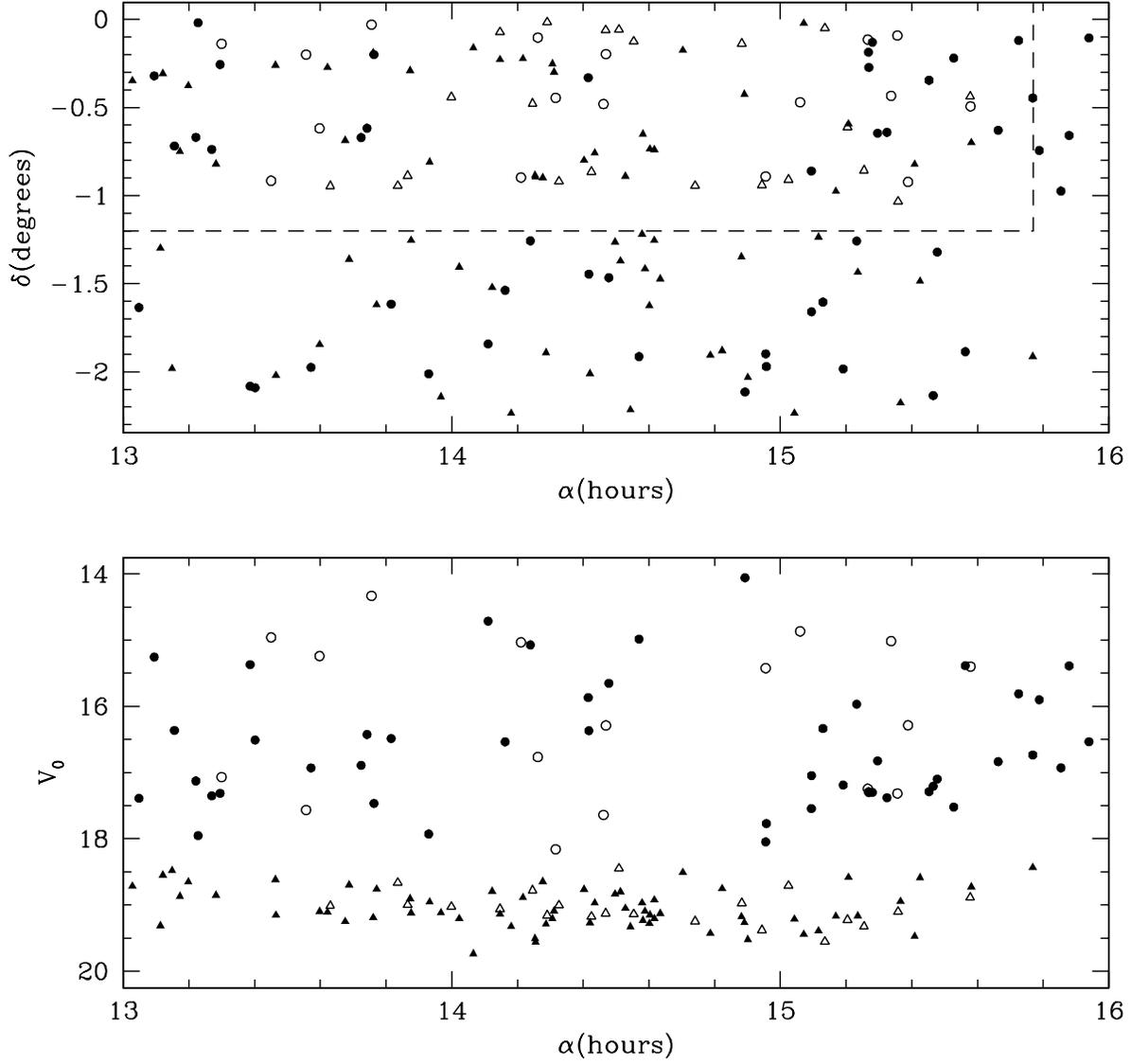}
\caption{Top: distribution of the RR Lyrae stars in the sky. The
dashed line encloses the region of overlap with SDSS. Bottom:
distribution of the $V_0$ (extinction corrected) magnitudes of the stars
as a function of right
ascension. In both diagrams, open symbols are stars discovered first
by SDSS and recovered by our survey. Triangles are the stars in the clump
($V_0\ge 18.4$).
\label{fig2}}
\end{figure}

\clearpage
\begin{figure}
\plotone{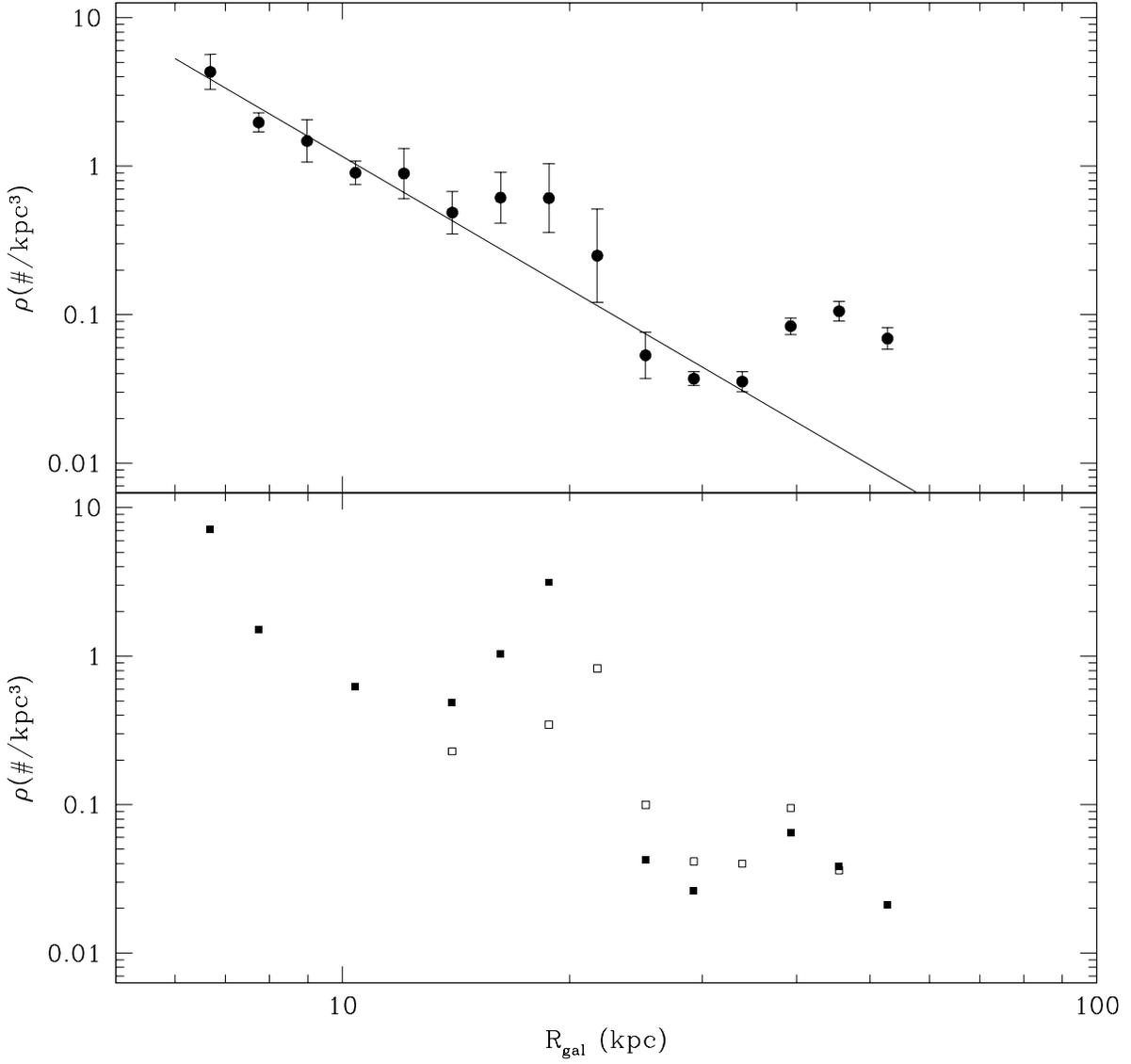}
\caption{(Top) Radial density profile of the RR Lyraes found in the
survey. (Bottom) Radial density profiles in the sub-region
$\alpha$=13.0-13.375$^h$ (open squares) and $\alpha$=15.250-15.625$^h$
(solid squares), which coincides with Pal 5.
\label{fig3}}
\end{figure}

\clearpage
\begin{figure}
\plotone{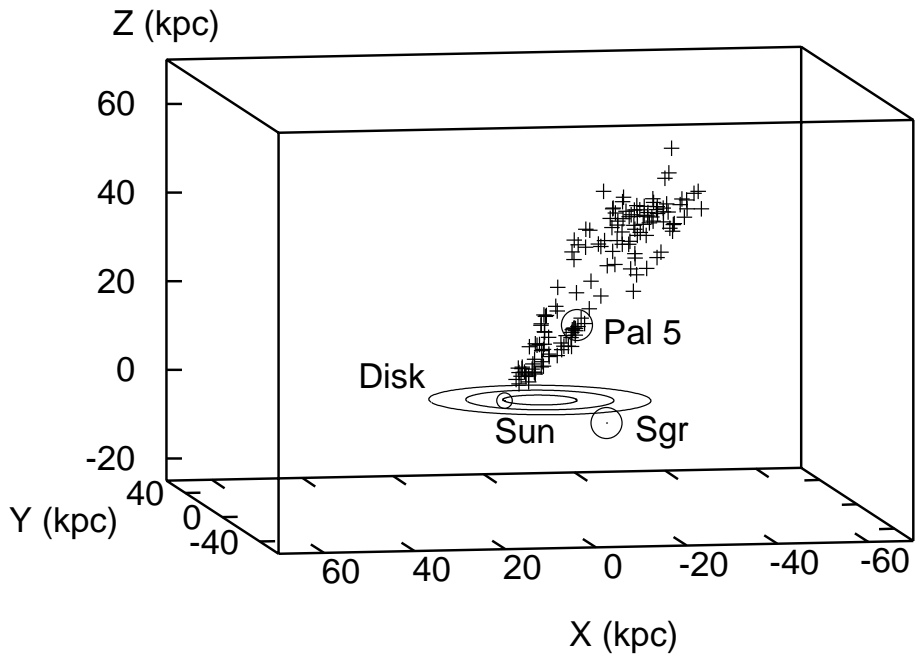}
\caption{3D view of the locations of the RRs (+'s), the Sun, the Sgr dSph
galaxy and the globular cluster Pal5. Three circles in the galactic
plane with radii of 8, 16 and 24 kpc indicate the solar radius and the
approximate extents of the optical and the HI disks, respectively.
\label{fig4}}
\end{figure}

\end{document}